\documentclass[prd,twocolumn,showpacs,preprintnumbers,amsmath,amssymb]{revtex4}
\vspace{-1cm}
\usepackage{amsmath,amssymb,graphicx}

\usepackage {amssymb}
\newcommand{\nc}{\newcommand}
\nc{\ba}{\begin{eqnarray}}
\nc{\ea}{\end{eqnarray}}
\newcommand\be{\begin{equation}}

\newcommand\ee{\end{equation}}

\nc{\x}{{\bf{x}}}

\nc{\ka}{{\kappa_{10}  } }

\nc{\h}{  h_{w}}

\nc{\F}{ \bar F_{(3)}}

\nc{\C}{ C_{(2)\, MN}}
\nc{\rh}{ \rho_I^{1/4}}
\nc{\g}{ \bar g}

\input{epsf.sty}

\begin{document}


\title{ Energy Radiation by Cosmic Superstrings in  Brane Inflation}


\author{Hassan Firouzjahi}
\vspace{1cm}
\email{firouz@physics.mcgill.ca}


\affiliation{Physics Department, McGill University, 3600 University Street, Montreal, Canada, H3A 2T8 }

\begin{abstract}

The dominant method of energy loss by a loop of cosmic D-strings in models of warped brane inflation is studied. It is shown that  the energy loss via Ramond-Ramond field radiation can dominate by many orders of magnitude over the energy radiation via gravitational wave emission. 
The ratio of these two energy loss mechanisms depends on  the energy scale of inflation, the mass scale of string theory and  
whether it is a single-throat or a multi-throat inflationary scenario.
This can have important consequences for the detection of cosmic superstrings in the near future. 
It is argued that the bounds from cosmic microwave background
anisotropies and big bang nucleosynthesis  are the dominant cosmological sources to constrain the physical parameters of the network of cosmic superstrings, whereas the role of the gravitational wave-based experiments 
may be secondary.

\vspace{2cm}

Keywords : Cosmic Strings, Brane Inflation

\end{abstract}

\maketitle

\section{Introduction}
In models of brane inflation \cite{Dvali:1998pa}
cosmic superstrings are copiously produced \cite{Sarangi:2002yt} as either fundamental strings (F-strings), D1-branes (D-strings), or the bound state
of p F-strings and q D-strings( (p,q) strings).
In models of warped brane inflation, such as in \cite{Kachru:2003sx}, brane inflation takes place in a warped region of the Calabi-Yau (CY) compactification, the throat. 
There are several advantages in having a warped brane inflation. First,
the fine-tuning associated with the flatness of the inflationary potential may
be less severed compared to the original models of brane inflation. Second, in warped models one can have a number of anti-branes localized at the bottom of the inflationary throat while one or more mobile branes,  moving in the bulk or some other throats, get attracted towards the inflationary throat to initiate inflation. This may make the initial conditions for brane inflation more natural. The third advantage is that
the effective tension of cosmic superstrings located at the bottom of the inflationary throat can be significantly smaller than the naive expectation, $m_s^2$, where $m_s$ is the mass scale of string theory. This way the bounds on $G\mu$, the cosmic string tension measured in units of Newton constant $G$, from cosmic microwave background(CMB) and other observations \cite{Wyman:2005tu} can be easily satisfied \cite{Firouzjahi:2005dh}. For works on different aspects of cosmic superstrings see 
\cite{Copeland:2003bj, Firouzjahi:2006vp, Jackson:2004zg, Copeland:2006if, Dvali:2003zj}.

In the evolution of a  network of cosmic strings the power radiation by loops is
an important effect in driving the network towards the scaling regime.
In the case of conventional gauge strings large loops lose their energy via gravitational 
wave emission \cite{Vachaspati:1984gt}. On the other hand, in the case of 
axionic string \cite{Davis:1985pt}
the dominant source of energy loss by loops is via scalar Boson radiation\cite{Vilenkin:1986ku}.

As observed in \cite{Dvali:2003zj}, there are interesting similarities and differences between an axionic string and a D-string.
An axionic string couples to an anti-symmetric two-form potential which in four dimensions is the Hodge dual of a scalar, the axion. Similarly, the D-string couples to the Ramond-Ramond (RR) anti-symmetric two-form potential $C_{(2)}$. 
Axionic strings emit long-range scalar Bosons, whereas the D-strings radiate long range 
massless RR-fields. The differences are also intriguing. Unlike axionic strings, D-strings in flat space-time are gravitationally suppressed, as we shall see later. 

Since the D-strings emit massless RR-particles, one may wonder what is the dominant source of energy loss by a cosmic D-string loop. From the fact that both the
zero modes of the graviton and RR-field belong to the massless sector of the underlying ten-dimensional closed string theory, one may expect that the power radiation
via RR-particles emission is comparable to the gravitational wave emission. We shall see that in a flat 
background compactification  this is more or less true. On the other hand,
in the presence of a warped geometry the situation becomes non-trivial. The key points are: a)- The warp factor couples to the Dirac-Born-Infeld (DBI ) part of the D-string action, while it does not show up in the Chern-Simons part of the action. The effective tension is lowered by two powers of the warp factor, whereas the charge remains intact as in flat space-time.
b)- In the presence of a warped geometry, and upon dimensionally reducing the theory to four dimensions, the normalization of the RR-field zero mode is significantly different from the normalization of the graviton zero mode. This was noticed in \cite{Mukhopadhyaya:2002jn, Mukhopadhyaya:2007jn} 
and plays a crucial role in our analysis here.

In this paper we compare the energy loss via RR-field radiation to that of gravity wave emission by a loop of cosmic D-string.
The outline of the paper is as follows. In section 2 we present the effective four-dimensional action for a D-string in a warped compactification. In section 3
we study the effect of different warp factors on the RR-field zero mode normalization.
In section 4 the power radiations by the loop in a single-throat and two-throat inflationary scenarios are calculated. The constraints from Big Bang Nucleosynthesis (BBN) are briefly studied in section 5 followed by the conclusions.



\section{The Action}
Here, we obtain the four-dimensional action for a straight infinite cosmic D-string extended along the z-direction, where the D-string sources the RR anti-symmetric
two forms potential $C_{(2)\, MN}$. Here and in the following, capital indices $M, N,...$ are the ten dimensional indices, while the Greek indices
stands for the four-dimensional Minkowski coordinates.

We start with the ten-dimensional {\bf IIB} string theory action, containing gravity and the action for $C_{(2)}$ form field
\ba
\label{action1}
S_{IIB} = \frac{1}{2\, \ka^{2}} \int d^{10} x \, \sqrt{- G }  \left[ R- \frac{g_{s}}{12} F_{(3)}^{2}  
\right] + S_{local}
\ea
where $g_{s}$ is the string coupling 
and $G_{MN}$ and $R$ are the  ten-dimensional metric and Ricci scalar, respectively. 
The ten-dimensional gravitational constant is $\kappa_{10}^2=(2\pi \sqrt{\alpha'} )^8/ 4\pi= m_s^{-8}$, where $m_{s}$ is defined as the  string theory mass scale.
The anti-symmetric  RR three-form strength, $F_{(3)}$, is constructed from $C_{(2)}$ 
\ba
F_{(3) \, MNP} = \partial_{M} \, C_{(2)\, NP} + \partial_{N} \, C_{(2)\, PM}+ \partial_{P} \, C_{(2)\, MN} \, .
\ea
There are other terms in the action containing axion-dilaton and various form fields and potentials, but we assume that the axion-dilaton is fixed 
as in \cite{Giddings:2001yu} and forms other than $C_{(2)}$ are set to zero.

The local parts of the action from a D-string  contains DBI and Chern-Simons terms 
\ba
\label{loc1}
S_{local} = -\mu_{1} g_{s}^{-1} \,  \int d \, t \, d \, z \sqrt{-|\gamma_{ab} |} 
+ \, \mu_{1}  \,  \int d \, t \, d \, z \, C_{(2) t\, z}
 \ea
where $\mu_{1}= 1/(2 \pi \alpha')$ is the string charge and $\gamma_{ab}$
is the metric induced on the string world-volume
\ba
\gamma_{ab} = G_{MN} \, \partial_{a} X^{M} \, \partial_{b} X^{N} \quad  , \quad
\{a, b\} = \{t, z\} \, .
\ea

As explained before we are interested in warped brane inflation where the background metric is in the form of warped geometry
\ba
\label{metric1}
ds^{2} =  \h^{2}(y) \, dx^{\mu} dx_{\mu} + g_{mn}(y) d\,y^{m}  d\, y^{n}
\ea
where $y^{m}, y^{n},...$ represent the internal six-dimensional CY coordinates and the
warp factor $\h$ is only a function
of the internal coordinates, labeled collectively by $y$.

We will consider only the zero modes of the graviton and RR two-form potential and neglect their Kaluza-Klein (KK) modes.
This simplifies our analysis considerably, and the four-dimensional action is obtained by a direct dimensional reduction of the ten-dimensional action (\ref{action1}). The higher KK modes are massive and would not contribute to the long range force and their contribution in power radiation may be neglected in first approximation.
The zero mode of the graviton results in the conventional four-dimensional gravity. The zero mode
of  $C_{(2)}$, on the other hand, is the field that the cosmic D-string is charged under, 
as observed by a four-dimensional observer.

To calculate the contribution from the graviton zero mode suppose one perturbs the background metric (\ref{metric1}) such that $\eta_{\mu \nu}   \rightarrow \g_{\mu \nu} (x^{\alpha} ) $ , 
where $\g_{\mu \nu} $ is the metric used by the effective four-dimensional observer.
Decomposing the ten-dimensional Ricci scalar $R$ into its four-dimensional counterpart, $\bar R$, via 
$ R =\h^{-2}\,  \bar R +...$, the action for the gravitational zero mode, $S_{g}^{(0)}$, becomes
\ba
\label{Sg}
S_{g}^{(0)} &=& \frac{1}{2\, \ka^{2}} \int d^{4} x\,  \sqrt{-\g} \, \bar R \, 
\int  d^{6}y   \, \sqrt{ g_{(6)}}  \, \h^{2}(y) 
\nonumber\\
&=& \frac{M_{P}^2}{ 2} \int d^{4} x\,  \sqrt{- \g} \, \bar R  \, ,
\ea 
where $ g_{(6)}$ represents the determinant of the internal 
six-dimensional metric and 
\ba
\label{MP}
{M_{P}}^{2} = \frac{1}{ \ka^{2}} 
\int  d^{6}y   \, \sqrt{g_{(6)}}  \, \h^{2}(y) \, 
\ea
is the four-dimensional Planck mass, related to the Newton constant by
$ 8 \, \pi\, G= M_{P}^{-2}$.

To calculate the contribution from the zero mode of $C_{(2)}$, we note that $F_{(3)\, MNP}(x^{\mu})$ as observed by the four-dimensional observer 
has components only along the four-dimensional Minkowski coordinates and
\ba
F_{(3)}^{2}  &=& G^{\alpha \alpha'} G^{\beta \beta'} G^{\gamma \gamma'} 
F_{(3)\, \alpha \beta \gamma}  F_{(3)\, \alpha' \beta' \gamma'}  \nonumber\\
&=& \h ^{-6} \F^{2} \, .
\ea
where 
\ba
\F^{2}= \g^{\alpha \alpha'} \g^{\beta \beta'} \g^{\gamma \gamma'}   
F_{(3)\, \alpha \beta \gamma}  F_{(3)\, \alpha' \beta' \gamma'  }  \, .
\ea
Using this in (\ref{action1}), the action for the zero mode of $C_{(2)}$,
$  S_{C_{(2)}}^{(0)} $,  is 
\ba
\label{SC}
S_{C_{(2)}}^{(0)}  =  \frac{g_{s} }{24\, \ka^{2}}    \int d^{4} x\,  \sqrt{-\g}  \F^{2}\, 
\int  d^{6}y    \sqrt{ g_{(6)}}  \, \h^{-2}(y)
\ea

Comparing (\ref{Sg}) and (\ref{SC}) it is evident that in the flat background with no warping where $\h=1$ and neglecting the string coupling,  the zero modes of graviton and $C_2$ are both Planck-suppressed \cite{Dvali:2003zj}. 
This is not surprising since both the graviton and $C_{(2)}$ belongs to the massless sector of the original ten dimensional string theory. 
However, in the presence of warping the zero mode of $C_{(2)}$ and the graviton appears with
different normalizations \cite{Mukhopadhyaya:2002jn, Mukhopadhyaya:2007jn}. 
In order to account for this difference in couplings, we introduce the parameter $\beta$, defined by
\ba
\label{s}
\beta \equiv  \frac{ \int  d^{6}y   \, \sqrt{ g_{(6)}}  \, \h^{-2}(y)}{ \int  d^{6}y   \, \sqrt{ g_{(6)}}  \, \h^{2}(y)} \, .
\ea
Because of the negative powers of the warp factor in the numerator of (\ref{s}), $\beta$ is a sensitive
function of the warp factor. Physically, this means that the most warped regions of the compactification contribute the most for the normalization of $C_{(2)}$ zero mode. This is unlike the case of graviton zero mode where its normalization over the compactification is not sensitive to the warping. We shall see these differences explicitly in the following section.

The local part of action coming from the D-string , considering the effect of warping, is
\ba
\label{loc2}
S_{local} = -\mu_{eff}   \int d \, t \, d \, z \sqrt{-|\bar {\gamma}_{ab} |} 
+ \, \mu_{1}    \int d  t  d  z  C_{(2)  t z}
\ea
where $\bar{\gamma}_{ab} = \g_{\mu \nu} \, \partial_{a} X^{\mu} \partial_{b} X^{\nu}$ is the induced metric on the string
world-volume from four-dimensional metric,
\ba
\label{mueff}
\mu_{eff}= h_{I}^{2} \, \mu_{1} \, g_{s}^{-1} 
\ea
is the effective tension of the string and $h_{I}$ is the warp factor at the bottom of the inflationary throat.
The key point is that the warp factor couples to the DBI term, while it does not show up in Chern-Simons
part of the action. This is simply because the latter is a topological term, defined independent of metric. Also it is important to note that the
D-strings are produced at the end of the throat where the relevant warp factor is $h_I$.

Combining (\ref{Sg}), (\ref{SC}) and  (\ref{loc2}) the total action  for the zero modes of graviton and 
$C_{(2)}$ is
\ba
\label{S0}
S&=&  \frac{M_{P}^2}{ 2} \int d^{4} x\,  \sqrt{- \g} \,    \left(  \bar R - \frac{\beta\, g_{s}}{12 } \F^{2} \right )\nonumber\\
&-&\mu_{eff}    \int d \, t \, d \, z \sqrt{-|\bar{\gamma}_{ab} |} 
+  \mu_{1}    \int d \, t \, d \, z  C_{(2) t z} \, .
\ea

The geometry around an infinite cosmic string extended along the $z$-direction is \cite{Vilenkin:1981zs}
$ds^2= (\eta_{\mu \nu} - h_{\mu \nu}) \, dx^{\mu} dx^{\nu}$, where the non-zero components are 
\ba
\label{hxy}
h_{xx}=h_{yy}= 8\, G\, \mu_{eff}\,\ln(\frac{r}{r_0}) \, .
\ea 
Here $r$ is the polar radial coordinate and $r_0$ is a constant of integration. This is a flat geometry with the deficit angle 
$\Delta= 8\pi G \mu_{eff}$.
Similarly, we would like to find the solution for $C_{(2)}$ sourced by an infinite cosmic D-string. We work at the linearized 
level when $G\, \mu_1, G\, \mu_{eff}<<1.$ The equation of motion
for $C_{(2)}$ field is
\ba
\label{Ceq1}
\partial_{\alpha} F^{\, \alpha \beta \gamma}= \frac{-2 \, \mu_1}{\beta\, g_s\, M_P^2}  (\dot X^\beta X'^\gamma -\dot X^\gamma X'^\beta) \,    \delta(x,y) \, ,
\ea
where an overdot and a prime denote derivative with respect to $t$ and $z$ respectively and $X^{\alpha}$ represents the position of the D-string.  
Choosing the Lorentz gauge $\partial _\alpha \,  C_{(2)}^{ \, \alpha \beta}=0$, one obtains 
\ba
\label{Ceq2}
\nabla^2 C_{(2)\, t\, z} = \frac{\mu_1}{ \pi  \beta g_s M_P^2}  \,
\frac{\delta(r)}{r} \, ,
\ea
where $\nabla^2$ is the three-dimensional Laplacian. 
Eq. (\ref{Ceq2}) closely resembles the equation for gravitational field $h_{\mu \nu}$
and one finds
\ba
\label{Csol}
C_{(2)\, t z}= \frac{8\, G\, \mu_1}{\beta \, g_s} \ln(\frac{r}{r_0}) \, .
\ea
Combining this with (\ref{mueff}) and (\ref{hxy}) one finds the interesting result that 
\ba
\label{Ch}
C_{(2)\, t z }= \frac{h_{xx}}{ \beta \, h_I^{2}  } \, .
\ea

In a flat background where $h_I=\beta=1$, one finds that $C_{(2)\, t z }=h_{xx}$.
As explained before this is not surprising because in four dimensions both the graviton and $C_{(2)}$ zero modes are gravitationally suppressed and belong to the massless sector of the underlying ten-dimensional closed string theory.

Like in axionic string case, $C_{(2)}$ contributes to 
the string tension which diverges logarithmically. This divergence is  subject to cut-off
both at the scale of string core and at the largest scale corresponding to the current size of the Universe. Denoting the $C_{(2)}$ contribution to the string tension by $\mu_{RR}$, one finds
\ba
\label{muRR}
\mu_{RR} = \frac{\pi}{2} \beta \, g_{s} M_P^2  \, \int d\, r \, \F^2= 
 \frac{ \ln (d_0/\delta)  \, \mu_1^2}{ 2\pi\, \beta \,  g_s M_P^2} \, .
\ea
Here $d_0$ and $\delta$ correspond to the cut-off imposed 
from the current size of the Universe and the string core respectively and
one may take $\ln (d_0/\delta) \sim 100.$ 

Comparing $\mu_{RR}$ to the string effective tension is also 
instructive, giving
\ba
\label{muRReff}
\frac{\mu_{RR}}{\mu_{eff}} \simeq \frac{\ln (d_0/\delta) }{  \beta \, h_I^2  } 
\left( \frac{m_s}{M_P} \right)^2 \, .
\ea
In a flat background it is clear that the $C_{(2)}$ field correction to the tension is negligible because of the large gravitational suppression.

\section{Compactification}

In models of brane inflation, the inflation occurs inside the inflationary throat. The throat is a warped region of compactification which is smoothly glued to the rest of the CY manifold, the bulk, which is not warped. A particular well-studied model of the throat is the Klebanov-Strassler (KS) solution \cite{Klebanov:2000hb}. It consists of a warped deformed conifold where the infra-red (IR) region of the geometry is smoothly cut off. For many practical purposes, one may approximate the KS solution by an AdS solution
such that
\ba
ds^{2} = \frac{r^{2}}{L^{2}} dx^{\mu} dx_{\mu} + \frac{L^{2}}{r^{2}} dr^{2} + L^{2} ds_{5}^{2} \, ,
\ea
where $ds_{5}^{2}$ represents the five-dimensional base with the volume $v_{5}=16\, \pi^3/27$ and $L$ measures the curvature radius of the AdS space given by
\ba
\label{L}
L^{4}= \frac{27 }{4} \pi g_{s}\, N \, \alpha'^{2}  \, .
 \ea
Here $N$ is the number of fluxes used to create this background geometry.
To trust the low energy supergravity analysis $N$ is at the order
of $100-1000$. Also to keep the perturbative analysis under control, one
can take $g_s \sim 10^{-1}.$ Combined, a theoretically well-motivated construction requires $10 \lesssim g_s N \lesssim 100$.
The geometry is cut off at $r=r_{0}$ and the warp factor at the bottom of the throat is given by $h_{I} = r_0/L$. The magnitude of $h_{I}$  depends on the scale of inflation compared to the scale of string theory, $m_s$. One may take $h_I$ to be around $10^{-2}-10^{-3}$.

Besides the inflationary throat, the compactification may contain some other throats too. Specifically, there may exist a throat where the Standard Model of particles physics (SM) is confined on a brane(or anti-brane) located at
the bottom of the throat. The corresponding warp factor at the bottom of the SM-throat, $h_{SM}$, can be very small. For example, for $m_s$ at the order of GUT scale, $h_{SM}\sim 10^{-12}$. There are some advantages in
considering multi-throat scenarios. First, the hierarchy problem can be solved via Randall-Sundrum (RS) \cite{Randall:1999ee} mechanism independent of the scale of inflation. The second advantage is related to the stability of cosmic superstrings.
It is argued  \cite{Copeland:2003bj} 
that in the presence of a D3-brane( or anti-brane), cosmic superstrings
may dissolve to the brane world-volume and annihilate. In order to prevent this from happening, one natural choice is to put the SM-brane (or anti-brane)
in a throat separated from the inflationary-throat.

The normalization of the graviton and $C_{(2)}$ zero modes are controlled by $M_P$ and $\beta$, where the magnitude of these quantities depends on the compactification. In order to take the effect of compactification into account, we do as follows.
Denote the radius of the bulk by $R_{CY}$ such that $V_{bulk} \sim R_{CY}^{6}$.
Also denote the curvature radius and the 
size of each throat by $L_{\, i}$ and $R_{c}^{\,i}$ respectively, such that at $r=R_{c}^{\,i}$ the throat is glued to the bulk.
Consistency of the setup requires that $L_{ i} \sim R_{c}^{\, i} <R_{CY}$. From Eq. (\ref{MP}) one finds that
\ba
\label{MP2}
M_P^2 \simeq m_s^8 R_{CY}^6  \left[ 1+ \frac{8\pi^3}{27} 
\sum_i \left(\frac{L_{ i}}{R_{CY}} \right)^ 6
\right]
\simeq m_s^8 R_{CY}^6   ,
\ea
where the sum represents the contributions from the inflationary and the SM throats.
The last relation in (\ref{MP2}) holds for large enough compactification, $L_i<R_{CY}$,
which is physically well-motivated. More specifically, using Eq. (\ref{L}), in this limit one finds
\ba
\label{RCY}
\left(\frac{L_{\, i}}{R_{CY}} \right)^ 6 &\simeq &
\left(\frac{27 \, g_s N_i}{32\, \pi^{5/2}}\right)^{3/2} \, 
\left( \frac{m_s}{M_P} \right)^2  \nonumber\\
&\simeq& 10^{-2} \,  (g_s\, N_i)^{3/2} \,
\left( \frac{m_s}{M_P} \right)^2 \, .
\ea
Taking $g_s N\lesssim 100$, the condition $L_i<R_{CY}$ is equivalent
to $m_s \ll M_P$, which we assume is the case.

This is a hybrid of RS and the large extra dimensions \cite{ArkaniHamed:1998rs} scenarios. The physical mass scale at each throat is red-shifted compared to $m_s$ by the warp factor, but the largeness of $M_P$ is explained by the largeness of the compactification.
One can also consider the limit where a considerable volume of the compactification is warped, e.g. $L_I \sim R_{CY}$, but this is not physically well-motivated and we do not consider this possibility in the following.

Similarly, from Eq. (\ref{s}) one finds that
\ba
\label{s2}
\beta \simeq 1+  \left( \frac{3^{3/2}}{ 2^{9/2} \pi^{3/4} } \right)  \,  \left( \frac{m_s}{M_P} \right)^2\,
\sum_i h_i^{-2}  (g_s\, N_i)^{3/2} \,  ,
\ea
where as before the sum represents the contributions from the inflationary and the SM throat
with the corresponding warp factors $h_i$.

Comparing (\ref{MP2}) and (\ref{s2}) it is clear that $\beta$, unlike $M_P$, is a sensitive function of the warp factor.
A value of $\beta \sim 1$ indicates that the $C_{(2)}$ zero mode normalization has the same magnitude as that of the graviton zero mode. 
On the other hand, a value of $\beta$ significantly greater than unity indicates that
the coupling of $C_{(2)}$ zero mode to SM fields is strongly suppressed compared to that of graviton zero mode.

\section{Power Radiation}

As observed in \cite{Dvali:2003zj}, the cosmic D-string with the action (\ref{S0}) closely resembles the
axionic string, where $C_{(2)}$ plays the role of the two form potential for the axionic string. Like axionic strings which emit long range scalar bosons, the D-strings radiate massless RR particles with long range interactions. However, unlike the axionic string which is not gravitationally suppressed, the D-strings are gravitationally suppressed
as is evident from Eq (\ref{S0}). Therefore the dominant method of energy loss for an oscillation loop of axionic string is via boson radiation. Naturally, one may wonder what is the dominant channel of energy loss  for loops of cosmic D-strings.
Here we would like to answer this question with special 
attention to the effects of the warp factor.

The analysis of energy loss via boson radiation was studied in \cite{ Davis:1985pt,  Vilenkin:1986ku}. 
With the replacement $\C \rightarrow  \frac{1}{2} (\beta g_{s} \, M_{P} )^{1/2} \,  \C $ the action 
(\ref{S0}) has the same form as that of the axionic string studied in \cite{ Vilenkin:1986ku} and we can formally borrow their
results. The RR power emission, $P_{RR}$, therefore is
\ba
\label{powerC}
P_{RR}=  \frac{  \Gamma_{RR}\, \mu_{1}^ 2}{  4\pi^2 g_{s} \beta M_P^2}
=\left( \frac{\Gamma_{RR} }{2\pi \ln (d_0/\delta) } \right) \, \mu_{RR} \, .
\ea
Here $\Gamma_{RR}$ is a numerical factor of order $\sim 50$.

On the other hand, the gravitational power emission, $P_{g}$  ,  from a loop of string with tension $\mu_{eff}$
is calculated to be $\Gamma_{g} \, G\, \mu_{eff}^2$ \cite{Vachaspati:1984gt}, where $\Gamma_{g}$ is another loop-dependent numerical factor of order $\sim 50$. Using this for our warped D-string example, one obtains
\ba
\label{powerg}
P_{g}= \Gamma_{g} \, G\,  ( h_I^2 \, \mu_{1} \, g_{s}^{-1}    ) ^{2} \, .
\ea

Comparing (\ref{powerC}) and (\ref{powerg}), we obtain 
\ba
\label{ratio}
\frac{P_{RR}}{P_{g} }= \left(\frac{2\, \Gamma_{RR} }{\pi\, \Gamma_{g} }\right) 
\frac{ g_{s}}{\beta\, h_I^4} \, .
\ea

In the limit where $g_{s}$ is very small and the warping effect is negligible,  $\beta \sim h_{I} \sim1$, the dominant source of power radiation is the gravitational one, as in the case of ordinary gauge strings. So far in the literature this channel was considered the standard method of energy loss for cosmic superstrings. However, in models of warped brane inflation where $h_{I}\ll1$ and $\beta$ is a sensitive function of compactification, the situation may be different as we shall see below.
A very low inflationary energy scale, corresponding to a very small 
$h_I$, tends to maximize the power radiation via RR channel. On the other hand, having a very small $h_i$ would increase $\beta$ like $h_i^{-2}$.
The competition between these two effects requires more careful consideration. In order to keep the discussions clear, we consider different cases separately.

\subsection{Single-Throat Scenario}

This is the case when the stability of cosmic superstring is not an issue
and we have a single throat compactification. 
Furthermore, if the energy scale of inflation is much higher than TeV, one should invoke mechanisms other than RS to explain the hierarchy problem.

Denote the energy scale of inflation by $\rho_I^{1/4}$ such that 
$\rh= h_I \, m_s$. Using this in Eq. (\ref{s2}) one obtains
\ba
\label{beta1}
\beta  &\simeq&  1+ \left( \frac{3^{3/2}}{ 2^{9/2} \pi^{3/4} } \right)  \, (g_s\, N_I)^{3/2}  \,
\left( \frac{m_s^2}{M_P\, \rho_I^{1/4}} \right)^2 \nonumber\\
 &\simeq&  1+ \frac{1}{10} \, (g_s\, N_I)^{3/2} 
\left( \frac{m_s^2}{M_P\, \rho_I^{1/4}} \right)^2 \, .
\ea

As mentioned earlier, a  theoretically well-motivated construction implies 
that $g_s N_I \lesssim 100$. So in the following the pre-factor for the 
second term in the right hand side of Eq. (\ref{beta1}) is $\lesssim 100$.

For a given value of $m_{s}$, and depending on the scale of 
inflation, $\rh$, two limits are distinguished here.

\subsubsection{ High-Scale Inflation }

This limit is defined when the second term in the right hand side of Eq. (\ref{beta1}) is considerably smaller than unity. 
For this to happen the energy scale of inflation is high enough such that
(in the sense of order of magnitude) 

\ba
\rh \gg \frac{m_s^2}{M_P} \, .
\ea 
In this limit $\beta \simeq 1$
and from Eq. (\ref{ratio}) one obtains
\ba
\label{high1}
\frac{P_{RR}}{P_{g} } \sim g_s \,  h_I^{-4} = 
g_s \, \left(\frac{m_s}{\rh}\right)^4 \, .
\ea

This can be significantly greater than unity. For example taking 
$h_I \sim 10^{-2}$ and $g_s \sim 10^{-1}$, one obtains 
$P_{RR}/P_g \sim 10^7$ and RR-particle radiation is
the dominant channel of energy loss for
the loops of cosmic D-strings.

Furthermore, using Eq. (\ref{muRReff}), for the ratio of $\mu_{RR}$ to $\mu_{eff}$ one obtains (assuming $ \ln (d_0/\delta) \sim 100 $ ) 
\ba
\frac{\mu_{RR} }{\mu_{eff}} \sim 10^2 
\left( \frac{m_s^2}{M_P\, \rho_I^{1/4}} \right)^2 \, ,
\ea
which is significantly less that unity as defined by our limit here.

This limit can be also continued to the case when the scale of inflation is
such that the second term in the right hand side of Eq. (\ref{beta1}) is comparable to unity. In this case $\beta \sim 1$, 
$\rh > \frac{m_s^2}{M_P}$ and Eq. (\ref{high1}) still applies. 
The dominant source of energy loss is still via RR-field radiation, 
however now $\mu_{RR} \sim \mu_{eff}$.

\subsubsection{Low-Scale Inflation}

This limit is the opposite of the previous limit where the scale of inflation is low enough such that(in the sense of order of magnitude)
\ba
\rh \ll \frac{m_s^2}{M_P} \,,
\ea
and the second term in Eq. (\ref{s2}) is considerably bigger than the first term. In this limit one finds
\ba
\frac{P_{RR}}{P_{g} } \sim 10\, g_s \,    (g_{s} N_{I})^{-3/2}  \left( \frac{M_P}{\rh} \right)^2 \, .
\ea
This is also significantly bigger than one. For example, taking 
$g_{s} \sim 10^{-1}$, $g_s N_{I} \sim 10^{2}$ and $\rh \sim 10^9 GeV$, one finds that $P_{RR}/P_g \sim 10^{15}$.

Furthermore
\ba
\label{lowscale}
\frac{\mu_{RR} }{\mu_{eff}} \sim  10 \,   \ln (d_0/\delta)\, (g_{s} N_{I})^{-3/2}  \, .
\ea
For a reasonable value of $g_{s} N_{I}$ this implies $ \mu_{RR} \gtrsim \mu_{eff}$.


\subsection{Two-Throat Scenario}

As explained previously, the advantage of the two-throat scenario is that the cosmic superstrings
produced in the inflationary throat are stable(in the absence of extra branes or anti branes);
furthermore  the energy scales of inflation and SM physics are separated.

For comparable value of $g_{s}N_{I}$ and $g_{s}N_{SM}$ and assuming $h_{SM}<<h_{I}$,
the SM throat contribution dominates in Eq. (\ref{s2}). Using the relation $TeV=h_{SM} m_{s}$, one obtains
\ba
\label{beta2}
\beta \simeq 1+ 10\,(g_{s} N_{SM})^{3/2}  \left( \frac{m_{s}}{10^{11} GeV} \right)^{4}
\ea

As in the case of single-throat compactification, two limits are distinguished here:


\subsubsection{ Low-Scale String Theory}
This is the limit where the scale of string theory is small enough such that $m_{s} \lesssim 10^{10} GeV$
which leads to
\ba
\frac{P_{RR}}{P_{g} } \sim g_s h_{I}^{-4} \, .
\ea
This is formally similar to the case of  high-scale inflation in single-throat compactification. With the same values for
$g_{s}$ and $h_{I}$ the energy loss via RR-field radiation is $10^{7}$ times bigger than the energy loss via gravity wave emission.
Furthermore, in this limit
\ba
\label{muratio}
\frac{\mu_{RR} }{\mu_{eff}} \sim \ln(\frac{d_{0}}{\delta})  ( \frac{m_s^2}{M_P \rho_I^{1/4}} )^2
\ll 10^{2} \frac{ (10^{10} GeV )^4 }{ (10^{21} GeV)^2  } \ll 1,
\ea
where in the last inequality it is assumed that $\rh$ is not bigger than 1 TeV.

\subsubsection{ High-Scale String Theory}

This limit is defined where the scale of string theory is high enough, 
$m_{s} \gtrsim 10^{11} GeV$, and the second term in (\ref{beta2}) dominates. This yields
\ba
\frac{P_{RR}}{P_{g} } \sim \frac{g_s}{10} (g_{s} N_{SM})^{-3/2} \,  \left( \frac{10^{11} GeV}{\rh} \right)^{4}
\ea
For a low enough inflationary scale, $\rh \lesssim 10^{9}GeV$, one finds that $P_{RR} >> P_{g}$.
Otherwise, $P_{RR} <P_{g}$ and the dominant mode of energy lose will be via gravitational radiation
as in conventional models of cosmic strings.

Finally, independent of the scale of inflation, one can easily check that $\mu_{RR}<< \mu_{eff}$ as in 
Eq. (\ref{muratio}).


\section{BBN Constraints }

In previous section it was shown that the energy loss via RR-particles radiation can dominate over energy loss via gravitational radiation by many orders of magnitude.
This can have significant consequences in searches  for cosmic superstrings in near-future observations. 
In particular, if the dominant source of energy loss is via RR-field radiation then constraints from BBN and CMB anisotropies are the dominant cosmological sources to impose bounds the parameters of the network of cosmic superstrings. On the other hand, experiments like LIGO or LISA designed for gravity wave detection and observations from pulsar timing
are secondary in search for cosmic superstrings.

An extensive energy loss by cosmic superstrings in the form of massless RR-particles alters the expansion rate of the Universe during BBN. This in turn changes the successful prediction for the abundance of light elements from standard BBN. This constraint usually is expressed in term  of bounds on adding an extra neutrino component, $\delta N_\nu$, in the energy budget of the Universe. Recent bounds from WMAP and the abundance of the light elements impose the constraint
$\delta N_\nu <0.85$ \cite{Cyburt:2004yc}(depending on how one fits the data, the bound can be even tighter). This leads to the conclusion that during BBN the fraction of energy density from RR-particles radiation is approximately less than $10\%$ of the total energy density from background radiation.
 
Here we would like to impose the BBN constraint on power radiation by cosmic superstrings.
As in the case of axionic strings, the evolution of the network of cosmic superstrings is not well-understood. To get a rough estimates on BBN constraints on power radiation we assume that the network of cosmic superstrings reaches a scaling regime like a network of ordinary gauge strings. This is a non-trivial assumption and a comprehensive treatment of the problem  must take
into account the differences between the evolution of a network of cosmic superstrings as compared to the evolution of a network of gauge strings.
In this regard our estimates here would be valid, at best, at the level of an order of magnitude. 

Our assumptions here are that the dominant method of energy loss is via RR-particle emission. 
Furthermore, $\mu_{RR}< \mu_{eff}$, which means that most of the energy of the loop comes from its warped tension. As we saw in the last section, both of these two assumptions are easily satisfied, except in the following cases: single-throat, low-scale inflation, where Eq. (\ref{lowscale}) leads to $\mu_{RR} \gtrsim \mu_{eff}$, and
two-throat, high-scale inflation with $\rh \gtrsim 10^{10} GeV$. 

Loops decaying  at time $t$ have length $\ell \sim P_{RR}\, t\, \mu_{eff}$. Define $n_l(\ell,t)\, d\, \ell$ as the number density of loops with length $\ell$ to $\ell+d\,\ell$ at the time $t$.
In the simplest approximation, one may take 
$n_\ell\sim \zeta\, \alpha^{1/2}\, (\ell \, t)^{-3/2}$, where $\zeta\sim 10$ is a numerical constant and $\alpha$ measures the size of loops compared to the Hubble radius at the time of loop formation.
The RR-radiation has discrete frequencies $\omega_n= 4 \pi n/ \ell$ and,
as in \cite{Vilenkin:1986ku}, suppose that most of the energy is transmitted by the first few harmonics, $n\sim 1$,  
with the energy density $\rho_{RR} \sim \mu_{eff} \, \ell \, n_\ell$. 
Then, using (\ref{mueff}) and (\ref{powerC}), the fraction of RR-particles energy density, $\rho_{RR}$, to the background energy density, $\rho_\gamma\sim \rho_c \sim 1/30 Gt^2$, is
\ba
\label{bbn1}
\frac{\rho_{RR}}{\rho_\gamma} \sim 30 \, \zeta 
\sqrt{ \frac{2\pi \alpha \ln(d_0/\delta) }{\Gamma_{RR} } }
\left( G \mu_{eff} \sqrt { \frac{\mu_{eff}}{\mu_{RR}} } \right) \, .
 \ea
Here $d_{0}$ represents the horizon size of the Universe at the time of BBN. Interestingly enough, only the combination in the above bracket is what fixed by Eq. (\ref{bbn1}).

The magnitude of the loop parameter $\alpha$ for gauge cosmic strings is under debate in recent literature \cite{Polchinski:2007qc}. 
Taking the large value for $\alpha$ suggested recently \cite{Vanchurin:2005pa}, $\alpha \sim 0.1$, and assuming $\rho_{RR}/\rho_\gamma < 1/10$, 
we obtain
\ba
\label{bbn2}
G\mu_{eff}  < 10^{-3} \left(\frac{\mu_{RR}}{\mu_{eff}} \right)^{1/2}\, .
\ea
This can be used to impose an order of magnitude constraint on the 
effective tension of the strings.
For example, applying this to case 1 of both single-throat and  two-throat scenarios in the previous section one obtains 
$G \mu_{eff} < 10^{-2} \, m_s^2/ M_P \, \rh$.

From our assumption $\mu_{eff}> \mu_{RR}$, so the lensing and the CMB anisotropies by cosmic strings are controlled by the effective tension. The recent bound on strings tension from CMB anisotropies is $G\mu_{eff} < 10^{-7}$ \cite{Wyman:2005tu}, 
which is consistent with Eq. (\ref{bbn2}). 
Probably Eq. (\ref{bbn2}) would be more useful if a lower bound on $G\mu_{eff}$ is known by some other mechanisms.

\section{conclusions}

In this paper the energy loss via RR-field radiation and gravity wave emission from a loop of cosmic D-strings are compared. It is shown that the former can take over the latter by many orders of magnitude. The warp factor couples to the Chern-Simons part of the D-string action, while it leaves the charge of the D-string intact.
This way the effective tension of the D-string is much smaller than the naive expectation, $m_s^2$, by two powers of warp factors. Furthermore, the normalization of the RR-field zero mode, denoted by $\beta$, is a sensitive function of the warped geometry. A large value of $\beta$ tends to cancel out the effect of warping. However, the decisive factor is the combination $\beta h_I^4$, and, as we demonstrated in section 4, this is typically much bigger than unity and makes the dominant channel of energy loss via RR-particles emission.

If the dominant method of energy loss by loops of cosmic D-strings is via RR-field radiation, then BBN and CMB anisotropies are the main cosmological sources to constrain the parameters of the D-string network. From CMB constraints one can impose bounds on the effective tension, $G\mu_{eff}$, while from BBN results one can constrain the ratio $\mu_{eff}/\mu_{RR}$. These results are model-dependent, namely whether one uses a single-throat or a multi-throat inflationary scenarios.

In our analysis, the contributions from stabilized dilaton and the KK modes of graviton, RR field and 
dilaton are neglected. A complete treatment of the power radiation by the loop of cosmic D-strings requires
that the decay via these massive particles also be included. However, If their masses are high enough, say 100 TeV or more, their contributions may be neglected at the first approximation. The question of power radiation via massive dilaton by strings loops was studied in 
\cite{Damour:1996pv}. It is shown that having a GUT-scale string tension consistent with BBN and other
cosmological constraints requires that the dilaton mass be no higher than 100 TeV.

To be specific, in our analysis we only considered the power radiation by cosmic D-strings. Our main results here are also applicable to power radiation by  F-strings. For this purpose, one replaces the RR two form potential, $C_{(2)}$, by the NSNS two form potential, $B_{(2)}$, and takes care of powers of $g_s$ accordingly. As long as warping and the normalization of the zero modes of the 
anti-symmetric tensor fields are concerned, an F-string is on the same ground as a D-string. This conclusion also generalizes to the case of (p,q) strings.


\vspace{1cm}

\begin{acknowledgments}{I would like to thank K.\ Dasgupta, J.\ Polchinski and A.\ Vilenkin for useful discussions and comments and A.\ Berndsen for comments on the draft.
This work is supported by NSERC. }\\
\end{acknowledgments}


\end{document}